# Deep learning for surgical instrument recognition and segmentation in robotic-assisted surgeries: a systematic review

**Fatimaelzahraa Ali Ahmed[1] · Mahmoud Yousef[2] · Mariam Ali Ahmed[3] · Hasan Omar Ali[2] · Anns Mahboob[2] · Hazrat Ali[6] · Zubair Shah[4] · Omar Aboumarzouk[1] · Abdulla Al Ansari[1] · Shidin Balakrishnan[1]**

## Abstract

Applying deep learning (DL) for annotating surgical instruments in robot-assisted minimally invasive surgeries (MIS) represents a significant advancement in surgical technology. This systematic review examines 48 studies that utilize advanced DL methods and architectures. These sophisticated DL models have shown notable improvements in the precision and efficiency of detecting and segmenting surgical tools. The enhanced capabilities of these models support various clinical applications, including real-time intraoperative guidance, comprehensive postoperative evaluations, and objective assessments of surgical skills. By accurately identifying and segmenting surgical instruments in video data, DL models provide detailed feedback to surgeons, thereby improving surgical outcomes and reducing complication risks. Furthermore, the application of DL in surgical education is transformative. The review underscores the significant impact of DL on improving the accuracy of skill assessments and the overall quality of surgical training programs. However, implementing DL in surgical tool detection and segmentation faces challenges, such as the need for large, accurately annotated datasets to train these models effectively. The manual annotation process is labor-intensive and time-consuming, posing a significant bottleneck. Future research should focus on automating the detection and segmentation process and enhancing the robustness of DL models against environmental variations. Expanding the application of DL models across various surgical specialties will be essential to fully realize this technology's potential. Integrating DL with other emerging technologies, such as augmented reality (AR), also offers promising opportunities to further enhance the precision and efficacy of surgical procedures.

## Abbreviations

| | |
|---|---|
| RAS | Robotic-assisted surgery |
| MIS | Minimally invasive surgery |

Extended author information available on the last page of the article

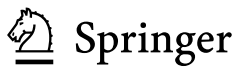

| | |
|---|---|
| AI | Artificial intelligence |
| DL | Deep learning |
| CNN | Convolutional neural network |
| FPS | Frames per second |
| CV | Computer vision |
| GAN | Generative adversarial network |
| HSV | Hue, saturation, value |
| FPN | Feature pyramid network |
| RPN | Region proposal network |
| ARAS | Augmented reality assisted surgery |
| PUMCH | Peking union medical college hospital |
| SGD | Stochastic gradient descent |
| mAP | Mean average precision |
| U-Net | U-shaped network |
| ResNet | Residual network |
| YOLO | You Only Look Once (computer vision model) |
| STswinCL | Swin transformer with joint space-time window shift scheme |
| IoU | Intersection over union |
| DiCE | Diverse counterfactual explanations |
| L1/L2 | Loss functions (refers to types of loss functions) |
| MATIS | Masked-attention transformers for instrument segmentation |
| SSD | Single Shot Detection |

# 1 Introduction

## 1.1 Role of deep learning in minimally invasive surgery

Robotic-assisted surgery (RAS) is a significant advancement in minimally invasive surgery (MIS) allowing surgeons to perform complex procedures using robotic arms, which reduces the need for an assistant surgeon. RAS requires small incisions for surgical tool insertion, leading to reduced blood loss and faster recovery times (Bramhe and Pathak 2022). Previous studies have evaluated various user-interfaces for controlling the movements of virtual minimally invasive surgical tools, which play a crucial role in enhancing the precision and usability of robotic systems in surgery (Shabir et al. 2022). Recent advancements in artificial intelligence (AI), particularly deep learning (DL) algorithms, offer immense potential to revolutionize surgical training and outcomes in MIS. The widespread implementation of RAS has significantly advanced MIS, enhancing surgical precision and instrument control. As of 2023, the global adoption of RAS systems has reached a remarkable milestone, with 7,733 units installed, paving the way for over 10 million robotic surgeries spanning various surgical disciplines such as general surgery, urology, gynecology, and cardiothoracic surgery (Peng et al. 2023). This proliferation of RAS has generated a vast amount of video data, presenting an untapped potential for training DL models to capture essential aspects of these surgeries. Figure 1 below, adapted from the 2024 earnings report of Intuitive Surgical Operations Inc., manufacturer of the Da Vinci Robotic systems that are widely used globally, shows the worldwide procedure trend for RAS from 2018 to 2023, highlighting a 22% growth in 2023, with clear increase in usage across Urology, Gynecology, and General Surgery disciplines (Intuitive Surgical Inc. 2024).

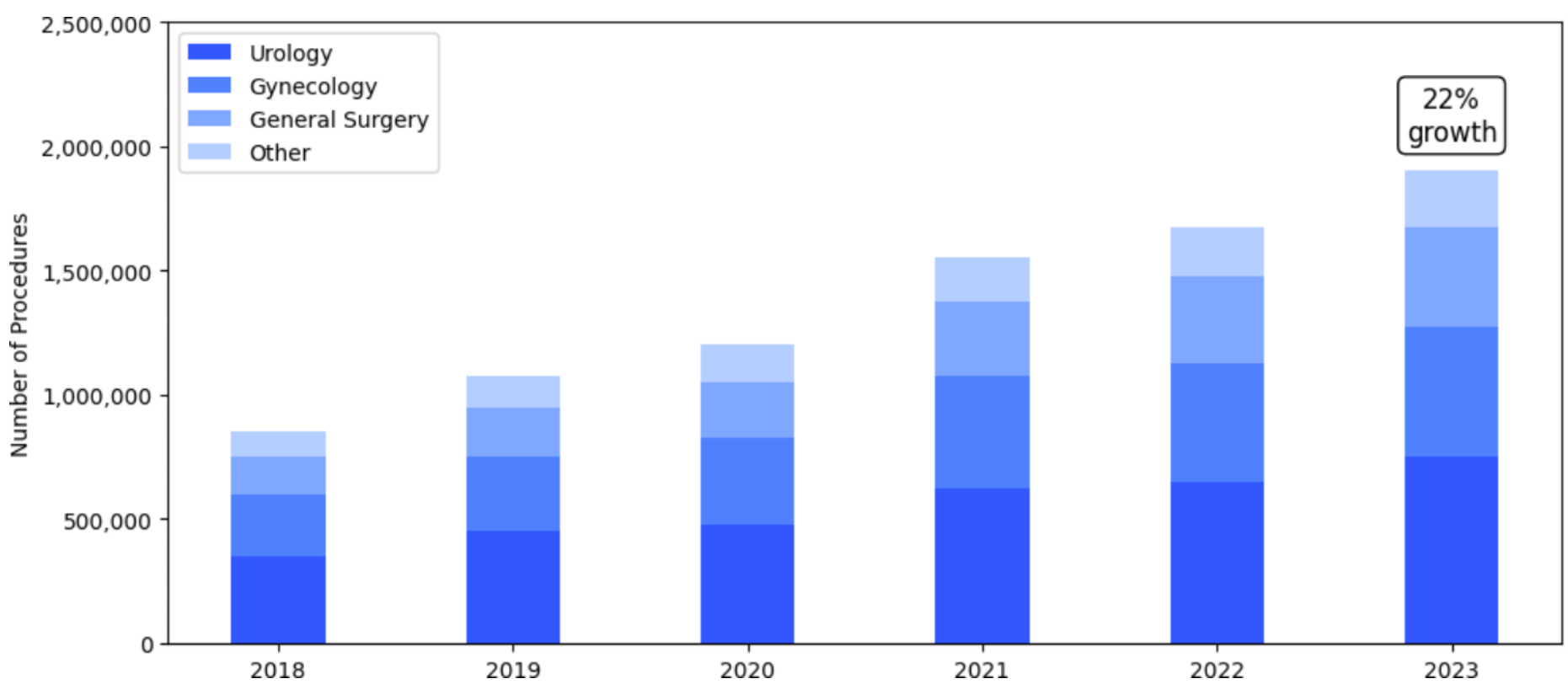

**Fig. 1** Annual global trend of RAS procedures using DaVinci robotic surgery systems, adapted from 2024 earnings reported by Intuitive Surgical (Intuitive Surgical Inc., 2024)

### 1.2 Current challenges in instrument recognition, segmentation, and dataset generation

The application of deep learning models in medical imaging and surgical procedures has demonstrated significant improvements in accuracy and efficiency compared to traditional image processing techniques and manual methods which often involve time-consuming, labor-intensive processes and may lack the precision that deep learning models can provide, particularly in tasks such as tool detection and segmentation (Ansari et al. 2022) and (Dakua et al. 2019)and (Yusuf et al. 2022). In-video instrument recognition and segmentation are crucial for understanding surgical processes, providing insights into the surgical phases and activities, and aiding the objective evaluation of a surgeon's skill and technical competency (Birkmeyer et al. 2013) and (Scally et al. 2016). These capabilities are vital for enhancing surgical training and potentially improving patient outcomes. Studies have already demonstrated the feasibility and effectiveness of instrument recognition across various surgeries, significantly impacting surgical training and competency evaluation (Kawka et al. 2021). However, a significant challenge remains in using DL for automated instrument recognition: the scarcity of correctly labeled, representative data.

A high-quality annotated dataset creation is a multi-step process, as depicted in Fig. 2 below. Extracted surgical videos need to be anonymized to remove all patient identifiers or frames revealing any personal details (e.g. faces), and appropriately down sampled to reduce overlap between frames without affecting data quality. For an average surgical procedure lasting between 60 and 90 min recorded at 24 frames per second (FPS) the total number of frames would be anywhere between 50,000 and 100,000 after down sampling. Subsequently, these frames need to be annotated using various commercially available software for various forms of segmentation or detection of the surgical tools, which then needs to be cross verified by expert surgeons. This process generates a high-quality curated dataset, which can then be utilized for training a DL model (De Backer et al. 2022).

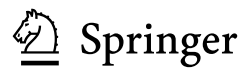

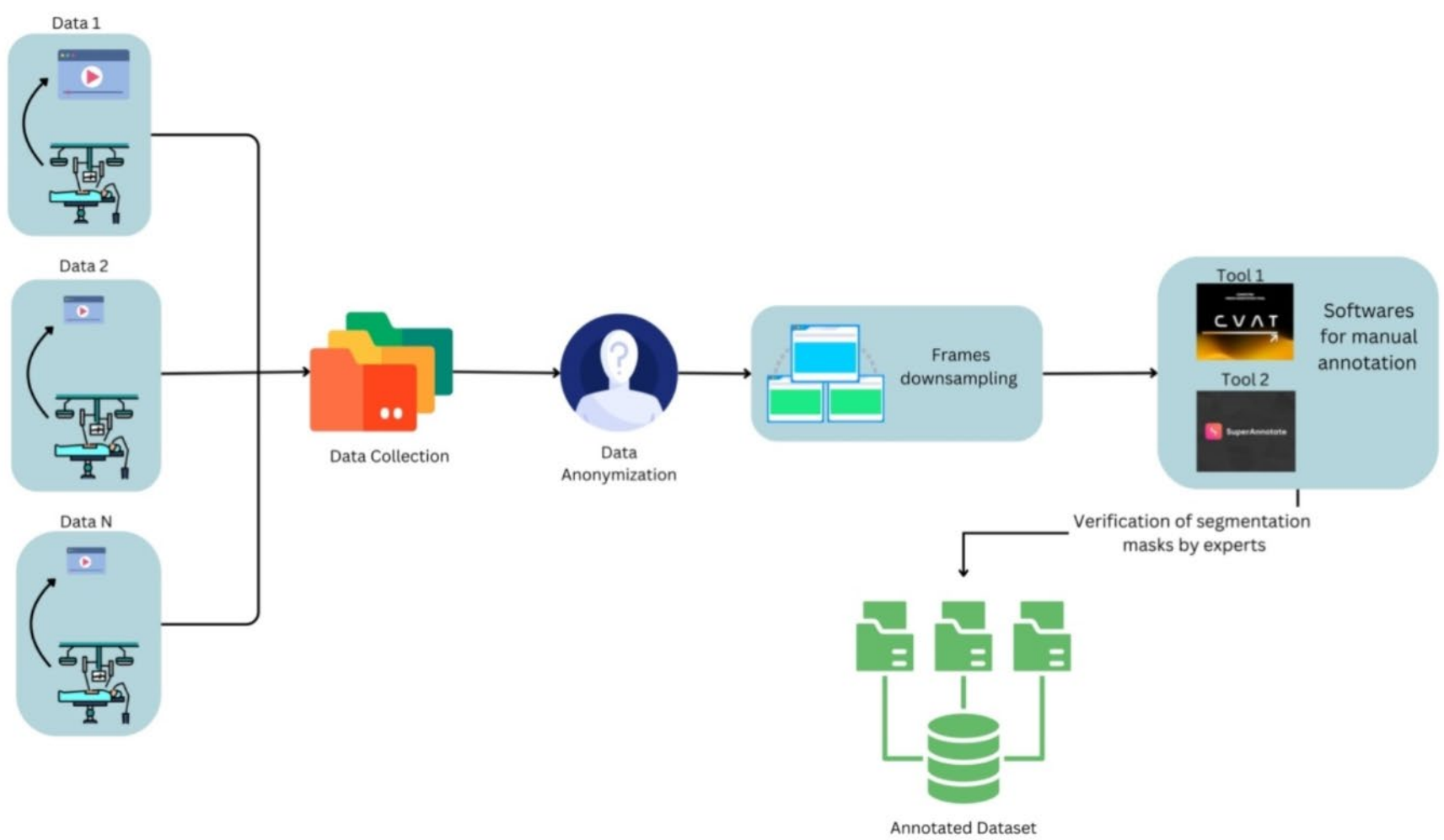

**Fig. 2** Multi-step process for high quality annotated dataset creation

Clearly, creating large training datasets through the manual process of human labeling is laborious and time-consuming, posing a significant challenge for the practical implementation of DL techniques for surgical tool detection and segmentation. The arduous and time-intensive task of manual annotation also demands the scarce and valuable time of expert surgeons, taking time away from their service in providing healthcare delivery. This highlights the need for automated, accurate, and efficient segmentation methods. The clinical necessity for a solution is clear: leveraging the abundance of raw RAS videos to train DL models for precise detection and segmentation of surgical tools, that can enhance the quality of surgical training videos and tile the way for advanced surgical analytics and automation, without compromising the valuable time of surgeons and creating a burden on healthcare delivery resources. Data annotation is essential to train any DL model for certain surgical applications. Based on the application complexity the images are processed accordingly.

Images are processed using various image processing techniques, such as resizing, filtering, and normalization, based on the model's specifications. The specific use case of the dataset determines the computer vision (CV) technique to be employed, as illustrated in Fig. 3 below. The most basic CV method is classification shown in Fig. 3b belw, which is generally unsuitable for surgical use unless combined with localization, resulting in object detection (Fig. 3c). Binary segmentation is another technique, where the frame is partitioned into two components: the object of interest and the background (Fig. 3d). This method creates a segmentation mask to delineate the object's exact boundaries without capturing detailed features (Fig. 3e). When multiple objects in the same frame are segmented, it is referred to as semantic segmentation (Fig. 3f). Table 1 below provides a detailed explanation of different annotations.

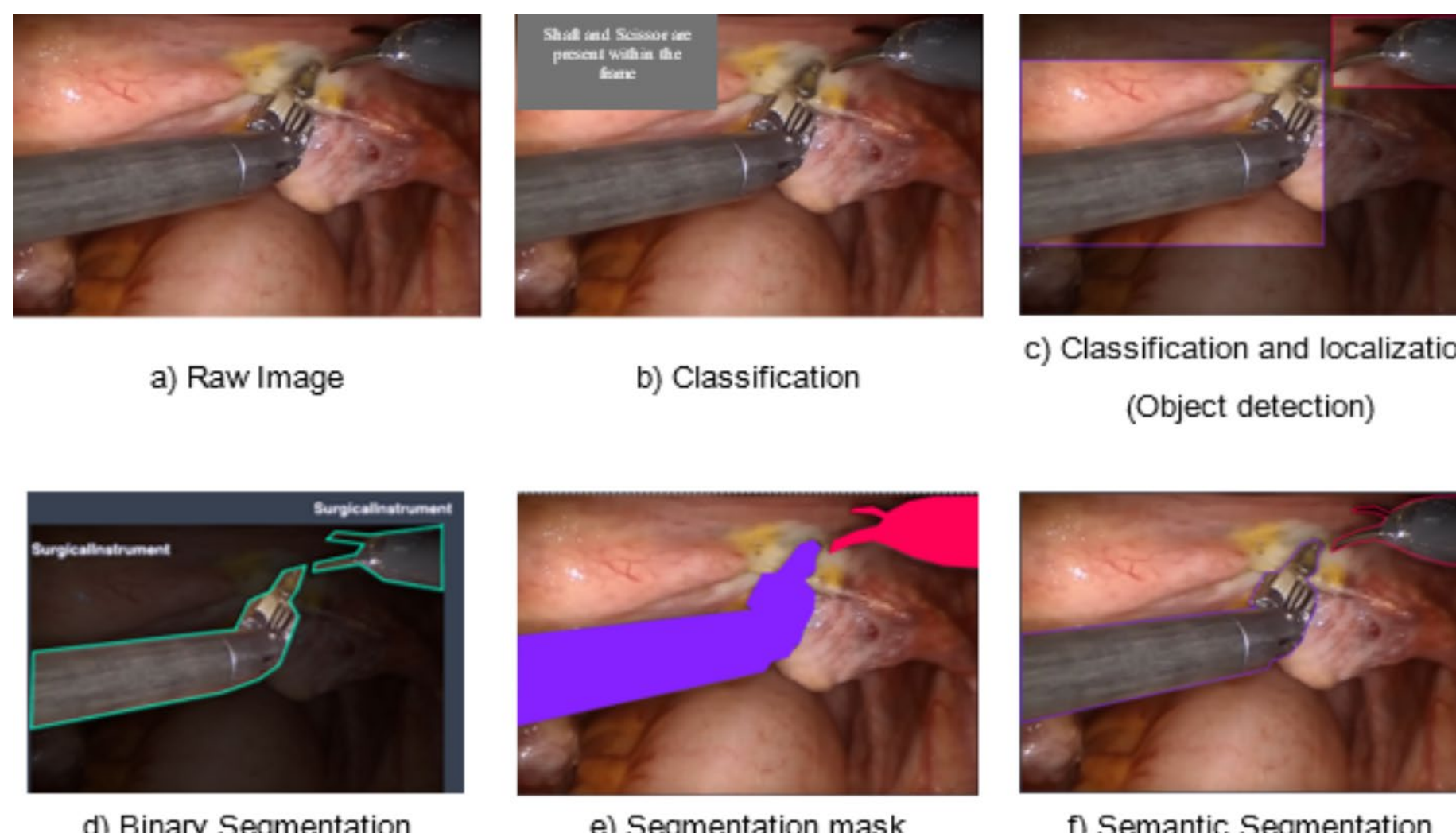

**Fig. 3** Different computer vision methods for detection of surgical tools

**Table 1** Definitions of different subclasses of image segmentation and classification, adapted from (IBM 2023)

| Subclass | Definition |
|---|---|
| *Image segmentation: technique for dividing an image into meaningful parts or segments to facilitate analysis* | |
| Semantic segmentation | Assigns a class label to each pixel in an image, allowing for a comprehensive understanding of the scene at a pixel level without differentiating between individual objects of the same class |
| Binary segmentation | Dividing an image into two distinct regions or classes. Basically, separating the foreground (objects of interest) from the background. The result is a binary image where pixels are assigned one of two values, commonly 0 (representing the background) and 1 (representing the foreground) |
| *Image classification: methods to categorize data into predefined classes or categories* | |
| Binary classification | A type of classification where the model divides the data into two distinct groups. It is used when there are only two possible states, outcomes, or classes |
| Multi-class classification | Extends binary classification to scenarios where there are more than two classes. The model distinguishes between three or more classes rather than just two |

## 1.3 Limitations of current methodologies

Manually annotating datasets is a labor-intensive and time-consuming process, often requiring expert knowledge. These challenges, further compounded by the scarcity of accurately labeled data, small labeled-dataset sizes, and insufficient domain generalization, severely impede the training process of DL algorithms (Kitaguchi et al. 2022). To overcome this challenge, recent research has explored the development of automated and semi-automated annotation techniques, leveraging the power of DL models. These methods aim to reduce the dependency on manual annotation by utilizing existing data and creating synthetic datasets, thereby facilitating the generation of annotated datasets at scale. This systematic review investigates the application of various DL models and architectures, such as convolutional neural networks (CNNs), U-Net, and ResNet, in automating the annotation process for surgical instruments.

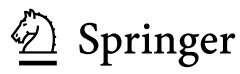

While these approaches show promise, they encounter obstacles such as variable lighting, visual obstructions, and the presence of extraneous objects (Lee et al. 2021) and (Kletz et al. 2019). Some techniques, such as the those explored by (Lee et al. 2021) highlight the use of DL networks like Faster R-CNN, Mask R-CNN, and SSD for instrument recognition with varying degrees of success. These methods, while promising, underscore the existing challenges of accurate instrument detection due to factors like the complexity of surgical scenes, similarity between different instruments, and dynamically fluctuating intraoperative environments.

### 1.4 Purpose of this review

Our motivation is thus to analyze current relevant studies that have successfully implemented DL models for the purpose of surgical instrument detection and segmentation. We aim to identify the strengths, limitations, and potential future directions for research in this domain. This evaluation is crucial for advancing the integration of DL in surgical practice, enhancing training, intraoperative guidance, and postoperative evaluation, ultimately improving patient outcomes. Thus, the primary objective of this review is to evaluate the effectiveness of these DL techniques in improving the accuracy and efficiency of surgical instrument detection and segmentation. The overarching aim is to guide stakeholders in identifying opportunities to improve DL capabilities to meet the stringent needs of contemporary surgical settings. Although a few previous evaluations have been published on the uses of AI in surgical video analytics, they either had a broad search scope, or they did not cover many contemporary studies in this niche. The comparison of our review with earlier published reviews on AI in robotic surgery is shown in Table 2 below.

## 2 Methods

This systematic review was conducted based on PRISMA guidelines (Page et al. 2021). Our systematic literature search was carried out across six databases: PubMed, Scopus, IEEE Xplore, Embase, Medline, and Web of Science. The primary search phrases encompassed three main topics: 'surgery', 'deep learning', and 'application'. The terms used in our DL study included not only the subject of DL itself, but also AI and ML, to ensure an extensive literature review and to avoid overlooking articles that utilize DL but are categorized under AI or ML, since DL is a subset of both. Though DL was presented to the ML community by (Dechter 1986), and modern DL era started in 2009, by Fei-Fei Li, who created ImageNet (Deng et al. 2009), DL was only introduced to the surgical annotation field for RAS videos in 2017, shown in Fig. 4. To ensure that we did not miss any early publications, our search looked at scholarly articles published from 2017 to 2024, with the aim of including all works that utilized DL in the context of RAS.

**Table 2** Comparison of our proposed review with published review articles in the same domain

| Publication | Key contributions | Comparison with this review |
|---|---|---|
| (Knudsen et al. 2024) | • The topic is broad as it included generic search terms for all AI techniques in robotic surgery.<br>• Search is limited to two years only i.e., from November 2021 to November 2023.<br>• Search is performed on a single (PubMed) database only. | • Our review focuses on deep learning techniques for robotic surgery.<br>• Our review covers studies during the last six years, 2017 to 2024.<br>• Our search is performed on six major databases. |
| (Zhang et al. 2024) | • The topic is broad as it included generic search terms for all AI techniques in robotic surgery.<br>• The review covers integration of AI with preoperative imaging and surgery.<br>• This is a narrative review with no formal structure as per PRISMA guidelines.<br>• Search strategy, inclusion/exclusion criteria or timeframe of review is not reported | • Our systematic review focuses on deep learning techniques for robotic surgery.<br>• Our review is not limited to preoperative procedures but covers preoperative and intraoperative integration of AI into robot-assisted surgeries.<br>• Our review follows the structured PRISMA guidelines for systematic review.<br>• Our review covers studies from 2017 to 2024. |
| (Amin et al. 2024) | • Narrative review which emphasizes how AI can be used to improve surgical outcomes and diagnostic accuracy in a variety of specialties, as well as how robotics and augmented reality can improve intraoperative performance and safety.<br>• Discusses the difficulties and moral dilemmas that come with incorporating AI into surgical practice. | • Our systematic review is focused more on the use of DL techniques in segmenting and classifying surgical instruments in robotic surgery. |
| (Moglia et al. 2021) | • The topic is broad as it covered search terms for all AI techniques in robotic surgery.<br>• Many recent developments are not covered as search was performed only until December 2020. | • Our review focuses on DL techniques for robotic surgery.<br>• Our review covers many recent and relevant studies beyond 2020. |
| (Ward et al. 2021) | • Application of computer vision in surgery, and how it can accurately identify operative phases (steps) and tools in surgical video. | • Our review covers many recent studies beyond 2020, specifically in DL |

The search strategy for each database included combinations of the following terms in their appropriate syntax:

- "Surgery" AND "deep learning".
- "Robot-assisted surgery" AND "artificial intelligence".
- "Minimally invasive surgery" AND "machine learning".
- "Surgical tool annotation" AND "deep learning".

A detailed breakdown of the search queries for each database can be found in the supplementary section under Appendix 1. Secondary filters were employed to include only English-language research articles, that specifically utilized DL models to label and annotate surgical processes, anatomy, and tools in minimally invasive robot assisted surgeries.

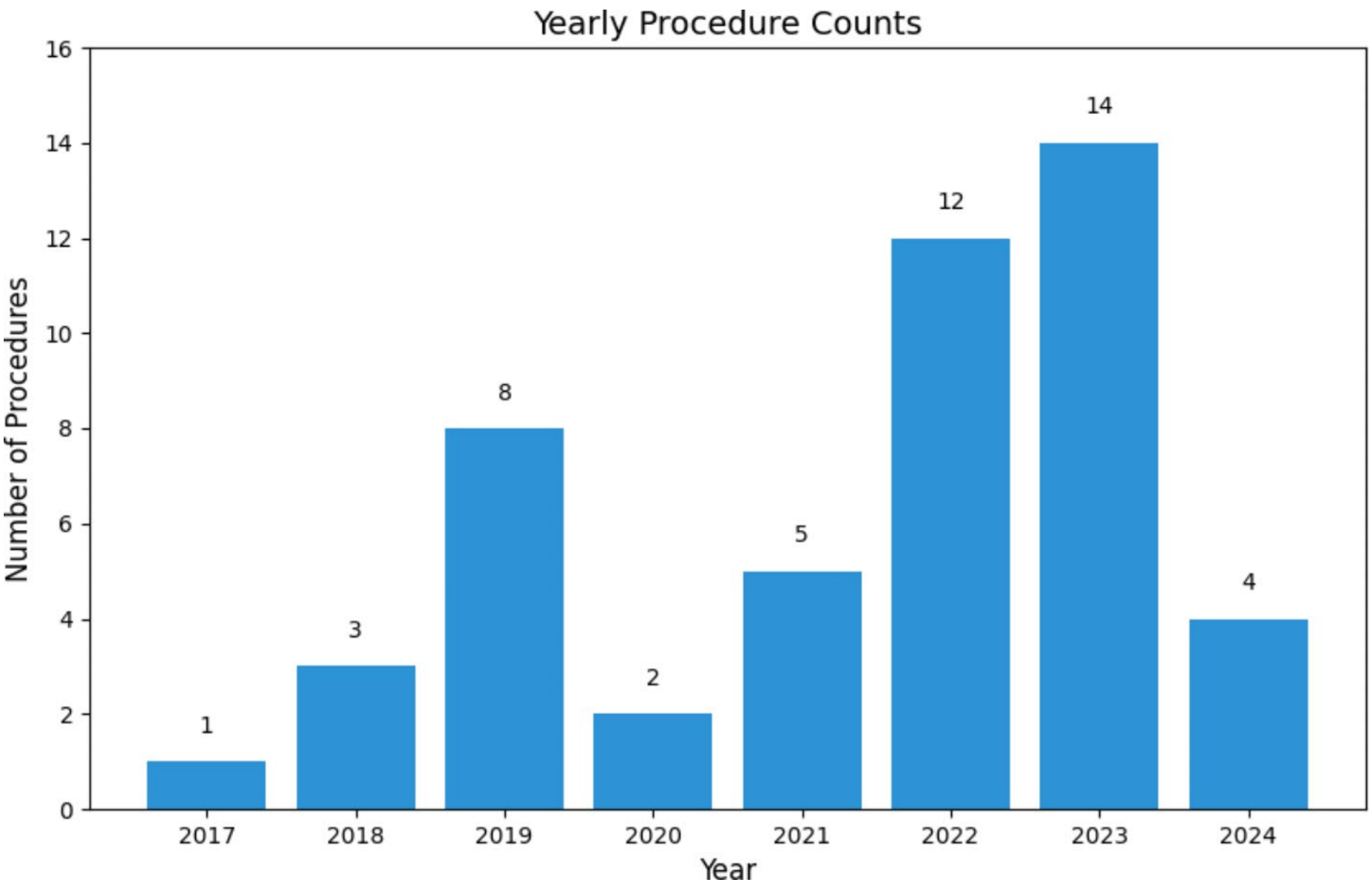

**Fig. 4** Number of published papers per year reporting on deep learning for robot assisted surgeries

## 2.1 Selection criteria

This review specifically focused on DL applications within MIS, due to their unique challenges in vision and tool manipulation. Within this context, we included papers that reported on all types of robot-assisted surgeries irrespective of the surgical sub-specialty, such as Nephrectomy and Prostatectomy (Urology), Hysterectomy (Gynecology), Sleeve-Gastrectomy (Bariatric Surgery), etc. We also included papers that reported on all types of DL algorithms, such as CNNs, GANs (Generative Adversarial Networks), and transformer. Papers reporting on open surgeries and minimally invasive procedures that were not robot-assisted were excluded. We also excluded papers that did not use DL techniques and instead reported on usage of AI or ML techniques. To ensure that our data originated from primary/original research offering the strongest direct evidence of DL's capabilities in this domain, we also excluded any review articles or meta-analyses. In addition, we eliminated publications of subpar quality that lacked rigorous methodology or adequate description, which would impede our ability to efficiently extract and validate data. Studies were excluded if they did not provide sufficient methodological detail to allow for replication or if they did not report on DL applications specifically related to surgical tool annotation. A PRISMA Checklist reporting applicable components of the systematic review standards is provided in Appendix 3 in the supplementary document.

The software 'Covidence' (Covidence 2024) was utilized for screening and selecting studies. Following the elimination of duplicate entries, the studies were evaluated by examining their titles and abstracts. Subsequently, the full-text versions of the selected studies were acquired and evaluated for potential inclusion in our review. The process of selecting studies was carried out by three authors working independently, and any disputes were resolved through discussion. If there was a lack of consensus, a fourth author was consulted.

## 2.2 Data extraction

Different data were extracted from the 48 included studies, which focus on annotating surgical tools in robot-assisted surgeries for in-vivo tissues using DL models. Our data extraction and analysis were performed using Microsoft Excel software that is part of the Office 365 suite (Microsoft 2024). Based on preliminary discussions with the surgeons at our institution, we extracted information from the selected articles that would be most important and beneficial to clinicians. The extracted data includes: (1) Title, year, and first author, (2) Purpose, (3) Limitations of the studies, (4) Deep Learning Model, (5) Annotation Method, (6) Clinical Applications, (7) Used Dataset, (8) Performance Metrics and Scores, (9) Network Architecture, (10) Number of Epochs and Batch, (11) Learning Rate, Optimizer and Loss Function, and (12) Used Hardware for Training. The detailed extracted data can be found in Tables 7 and 8 in Appendix 2 in the supplementary document.

## 2.3 Data synthesis and analysis

The extracted data were synthesized and analyzed based on key aspects of the studies that were identified by the authors as parameters of interest for this review. Based on each parameter, the extracted data were grouped into categories for further analysis. Such categorization helped in identifying patterns and trends within the data. The parameters and the sub-categorization for our data synthesis and analysis were as shown in Table 3 below:

Using the above parameters as a framework for data extraction and analysis, we were able to report our findings methodically ensuring that the synthesis provided clear insights into the application and performance of DL models in surgical tool detection and segmentation. Specifically, our methodical approach we used to aggregate and analyze the data from the included studies included the following components:

**Table 3** Key parameters and categories chosen as framework for data synthesis and analysis

| Parameter | Categories |
|---|---|
| Clinical use case | - Surgical workflow analysis<br>- Skill assessment<br>- Decision-making support<br>- Surgical navigation |
| Deep learning model | - U-Net<br>- ResNet<br>- CNNs<br>- Transformers |
| Annotation type | - Binary segmentation<br>- Multi-class segmentation<br>- Instrument part detection |
| Performance metrics | - Intersection over Union (IoU)<br>- Dice coefficient<br>- Accuracy<br>- Precision<br>- Recall |
| Data and dataset characteristics | - Types of datasets (public vs. private)<br>- Number of images<br>- Methods for data annotation (manual vs. automated) |
| Hardware and computational resources | - Specific GPUs or computational setups used for model training and inference |

- Data grouping: After categorizing the studies based on the parameters mentioned in Table 3, we systematically compared the findings across these categories to identify patterns, trends, and outliers. For example, we analyzed how different deep learning models performed in specific clinical use cases or how the choice of datasets impacted model accuracy.
- Performance comparison: We synthesized the performance data across studies, allowing us to draw conclusions about the relative effectiveness of different deep learning architectures in surgical tool detection and segmentation. This included comparative analysis of performance metrics like IoU and Dice scores across different studies.
- Insight generation: The synthesis primarily involved drawing insights from the aggregated data, such as identifying which deep learning models are most commonly used for certain types of annotation, or which models show the highest accuracy in specific surgical contexts. We also discussed the challenges and limitations observed across the studies, such as the need for large, annotated datasets and the dependency on high-quality data.

Through the aforementioned methodical approach towards data synthesis and analysis, we were able to present a comprehensive synthesis that not only highlights the current state of research but also provides actionable insights into the application and performance of deep learning models in the context of surgical tool detection and segmentation.

## 3 Results

A comprehensive systematic search identified a total of 10,472 studies. After eliminating duplicates and excluding studies based on title and abstract, a total of 1248 papers were selected for full-text screening. Of these, a total of 48 studies met our inclusion criteria and were included in the systematic review. Figure 5 below illustrates a PRISMA flowchart depicting the process of screening and selecting research.

The detailed extracted data is provided in the supplementary document in Appendix 2, as Tables 7 and 8. Table 7 provides a comprehensive overview of various included studies on the application of DL algorithms for annotating surgical instruments in robotic-assisted surgeries. It includes details on the study's title, purpose, limitations, type of annotation used, clinical use cases, and the specific DL algorithms employed, with each article listed with its corresponding year and reference for easy cross-referencing. The information in this table is expected to highlight the advancements and challenges in the field, emphasizing the impact of different DL models on surgical practice.

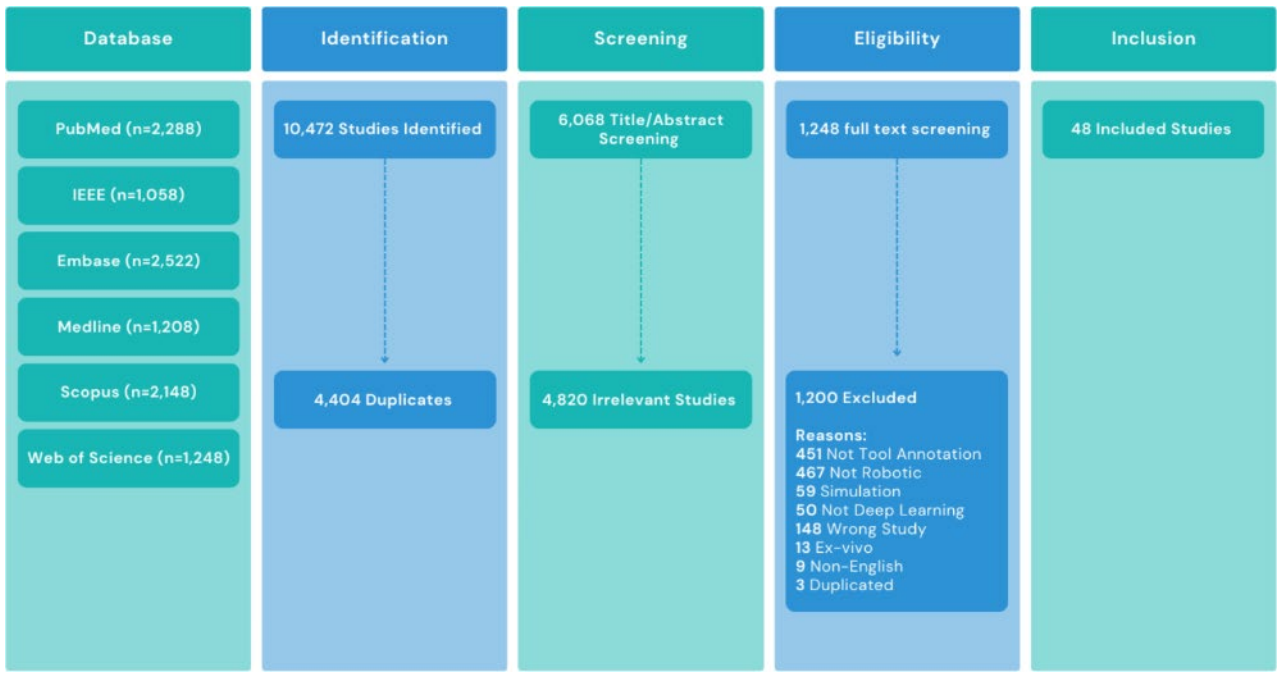

**Fig. 5** PRISMA flowchart depicting details of our study Selection process (Page et al. 2021)

Table 8 provides a detailed summary of various DL models used for surgical instrument detection and segmentation in robotic-assisted surgeries. We have included information on the model type, network architecture, batch size, number of epochs, learning rate, optimizer, loss function, performance metrics, and hardware used. Again, each study has been listed with its corresponding year and reference for easy cross-referencing. This comprehensive overview highlights the diversity in approaches and technical configurations used in the field, emphasizing the key parameters and outcomes that drive the effectiveness of these models in the included studies.

## 3.1 Study data

### 3.1.1 Annotation

The 48 studies included in this review utilized various annotation methods, including instrument object detection, classification with localization, segmentation or utilized the annotated data. Segmentation was categorized into semantic segmentation and binary segmentation. Semantic segmentation was employed in nine studies (Brandenburg et al. 2023, Ping et al. 2023, Zheng et al. 2022, Kletz et al. 2019, Hasan and Linte 2019, Kugener et al. 2022, Xia et al. 2023, Islam et al. 2019, Choi et al. 2021). The included studies reported detection and segmentation of various instruments such as scissors, graspers, forceps. etc. A detailed list of all the commonly annotated instruments is shown in Table 4 below. A variety of publicly available as well as private datasets were utilized to train the DL models to annotate surgical video frames automatically. These datasets are detailed in the subsection 3.1.2. For example, (Kletz et al. 2019) described a model capable of segmented and classifying 11 different instruments using distinct colors for each instrument in the frame.

It is also important to note that the included studies utilized different computer vision techniques for detection and segmentation of the instruments, as depicted in Fig. 6 below. Studies like (Hasan and Linte 2019), (Xia et al. 2023), (Islam et al. 2019), and (Lotfi et al. 2020) segmented different parts of a RAS instrument (such as tool-tip, shaft, etc.) and performed tool-tracking along with studies (Law et al. 2017) and (Yang et al. 2022). Notably, two studies successfully demonstrated the possibility of real-time semantic segmentation, including binary, part and multi-class segmentation (Xia et al. 2023) and (Islam et al. 2019). Additional applications included tool presence detection, segmentation, tool edge detection and tool mid-line detection, as presented by (Hasan et al. 2021). Tool tip detection was used in (Ping et al. 2023) and (Cai and Zhao 2020) while tool-joint detection was employed by (Law et al. 2017), (Du et al. 2018) and (Colleoni et al. 2019). The remaining studies performed binary segmentation or detection.

### 3.1.2 Datasets

All included studies have reported the use of specific datasets for training the developed DL models. The datasets were derived from different surgical procedures which are depicted in

**Table 4** Commonly used instruments that were annotated in the included papers

| Instrument | Papers reporting the instrument |
|---|---|
| Drop-in ultrasound probe 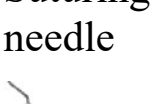 | (Yang et al. 2022), (Kalia et al. 2021), (Ni et al. 2020), (Xia et al. 2023), (Zinchenko and Song, 2021), (Ross et al. 2018), (Sestini et al. 2022), (Garcia-Peraza-Herrera et al. 2021), (Jin et al. 2019), (Ayobi et al. 2023), (Bian et al. 2023), (Colleoni and Stoyanov 2021), (Tukra et al. 2022), (Nema and Vachhani 2023), (Islam et al. 2019), (Lee et al. 2019), (F. Wang et al. 2023b, c), (Li et al. 2023), (J. H. Yang et al. 2022), (Hayoz et al. 2023), (Reiter 2022), (Suzuki et al. 2019), (Jin et al. 2022), (Ayobi et al. 2023), (Xu et al. 2022), (Tukra et al. 2022), (H. Wang et al. 2023b, c), (De Backer et al. 2022) |
| Suturing needle 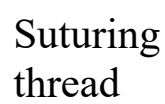 | (Jin et al. 2022), (Ayobi et al. 2023), (Xu et al. 2022), (Tukra et al. 2022), (H. Wang et al. 2023b, c), (De Backer et al. 2022) |
| Suturing thread 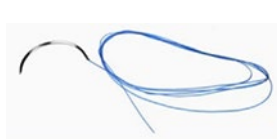 | (Jin et al. 2022), (Ayobi et al. 2023), (Xu et al. 2022), (Tukra et al. 2022), (H. Wang et al. 2023b, c) |
| Suction-irrigation device 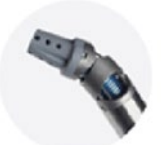 | (Jin et al. 2022), (Ayobi et al. 2023), (Xu et al. 2022), (Tukra et al. 2022), (H. Wang et al. 2023b, c), (De Backer et al. 2022) |
| Surgical clip 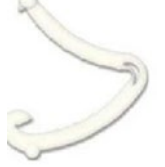 | (Jin et al. 2022), (Ayobi et al. 2023), (Xu et al. 2022), (Tukra et al. 2022), (H. Wang et al. 2023b, c)8/10/2024 9:27:00 AM |
| Large needle driver 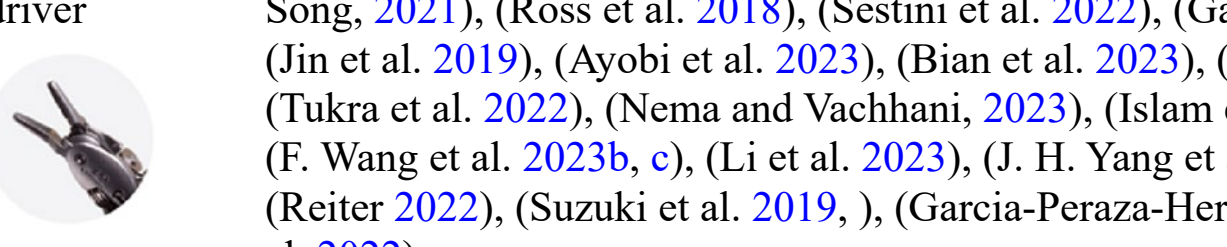 | (Yang et al. 2022), (Kalia et al. 2021), (Ni et al. 2020), (Xia et al. 2023), (Zinchenko and Song, 2021), (Ross et al. 2018), (Sestini et al. 2022), (Garcia-Peraza-Herrera et al. 2021), (Jin et al. 2019), (Ayobi et al. 2023), (Bian et al. 2023), (Colleoni and Stoyanov 2021), (Tukra et al. 2022), (Nema and Vachhani, 2023), (Islam et al. 2019), (Lee et al. 2019), (F. Wang et al. 2023b, c), (Li et al. 2023), (J. H. Yang et al. 2022), (Hayoz et al. 2023), (Reiter 2022), (Suzuki et al. 2019, ), (Garcia-Peraza-Herrera et al. 2021), (De Backer et al. 2022) |
| ProGrasp forcep 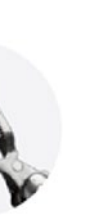 | (Yang et al. 2022), (Kalia et al. 2021), (Ni et al. 2020), (Xia et al. 2023), (Zinchenko and Song, 2021), (Ross et al. 2018), (Sestini et al. 2022), (Garcia-Peraza-Herrera et al. 2021), (Jin et al. 2019), (Ayobi et al. 2023), (Bian et al. 2023), (Colleoni and Stoyanov 2021), (Tukra et al. 2022), (Nema and Vachhani, 2023), (Islam et al. 2019), (Lee et al. 2019), (F. Wang et al. 2023b, c), (Li et al. 2023), (J. H. Yang et al. 2022), (Hayoz et al. 2023), (Reiter 2022), (Suzuki et al. 2019), (Garcia-Peraza-Herrera et al. 2021) |
| Monopolar curved scissor 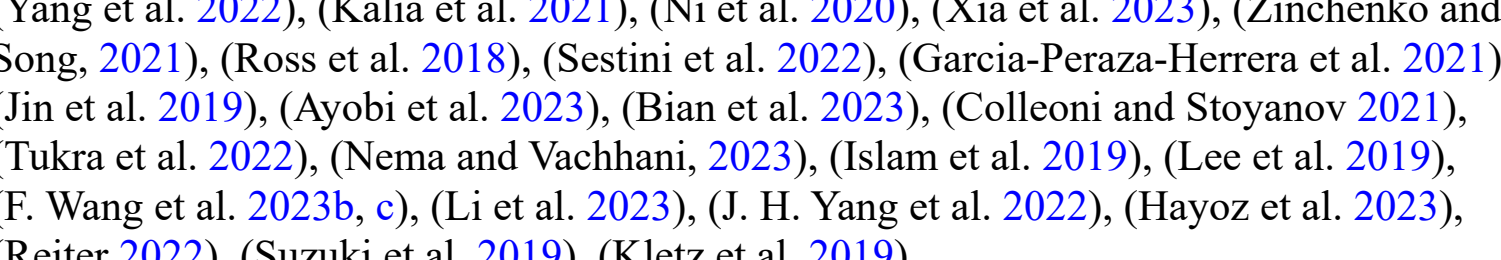 | (Yang et al. 2022), (Kalia et al. 2021), (Ni et al. 2020), (Xia et al. 2023), (Zinchenko and Song, 2021), (Ross et al. 2018), (Sestini et al. 2022), (Garcia-Peraza-Herrera et al. 2021), (Jin et al. 2019), (Ayobi et al. 2023), (Bian et al. 2023), (Colleoni and Stoyanov 2021), (Tukra et al. 2022), (Nema and Vachhani, 2023), (Islam et al. 2019), (Lee et al. 2019), (F. Wang et al. 2023b, c), (Li et al. 2023), (J. H. Yang et al. 2022), (Hayoz et al. 2023), (Reiter 2022), (Suzuki et al. 2019), (Kletz et al. 2019) |
| Grasper 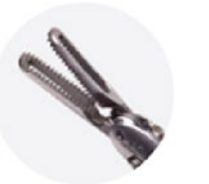 | (Yang et al. 2022), (Kalia et al. 2021), (Ni et al. 2020), (Xia et al. 2023), (Zinchenko and Song, 2021), (Ross et al. 2018), (Sestini et al. 2022), (Garcia-Peraza-Herrera et al. 2021), (Jin et al. 2019), (Ayobi et al. 2023), (Bian et al. 2023), (Colleoni and Stoyanov 2021), (Tukra et al. 2022), (Nema and Vachhani, 2023), (Islam et al. 2019), (Lee et al. 2019), (F. Wang et al. 2023b, c), (Li et al. 2023), (J. H. Yang et al. 2022), (Hayoz et al. 2023), (Reiter 2022), (Suzuki et al. 2019), (De Backer et al. 2022), (Kletz et al. 2019), (Garcia-Peraza-Herrera et al. 2021) |

**Table 4** (continued)

| Instrument | Papers reporting the instrument |
|---|---|
| Fenestrated bipolar forcep | (Yang et al. 2022), (Kalia et al. 2021), (Ni et al. 2020), (Xia et al. 2023), (Zinchenko and Song, 2021), (Ross et al. 2018), (Sestini et al. 2022), (Garcia-Peraza-Herrera et al. 2021), (Jin et al. 2019), (Ayobi et al. 2023), (Bian et al. 2023), (Colleoni and Stoyanov 2021), (Tukra et al. 2022), (Nema and Vachhani, 2023), (Islam et al. 2019), (Lee et al. 2019), (F. Wang et al. 2023b, c), (Li et al. 2023), (J. H. Yang et al. 2022), (Hayoz et al. 2023), (Reiter 2022), (Suzuki et al. 2019), (Kletz et al. 2019) |
| Vessel sealer | (Yang et al. 2022), (Kalia et al. 2021), (Ni et al. 2020), (Xia et al. 2023), (Zinchenko and Song, 2021), (Ross et al. 2018), (Sestini et al. 2022), (Garcia-Peraza-Herrera et al. 2021), (Jin et al. 2019), (Ayobi et al. 2023), (Bian et al. 2023), (Colleoni and Stoyanov 2021), (Tukra et al. 2022), (Nema and Vachhani, 2023), (Islam et al. 2019), (Lee et al. 2019), (F. Wang et al. 2023b, c), (Li et al. 2023), (J. H. Yang et al. 2022), (Hayoz et al. 2023), (Reiter 2022), (Suzuki et al. 2019) |

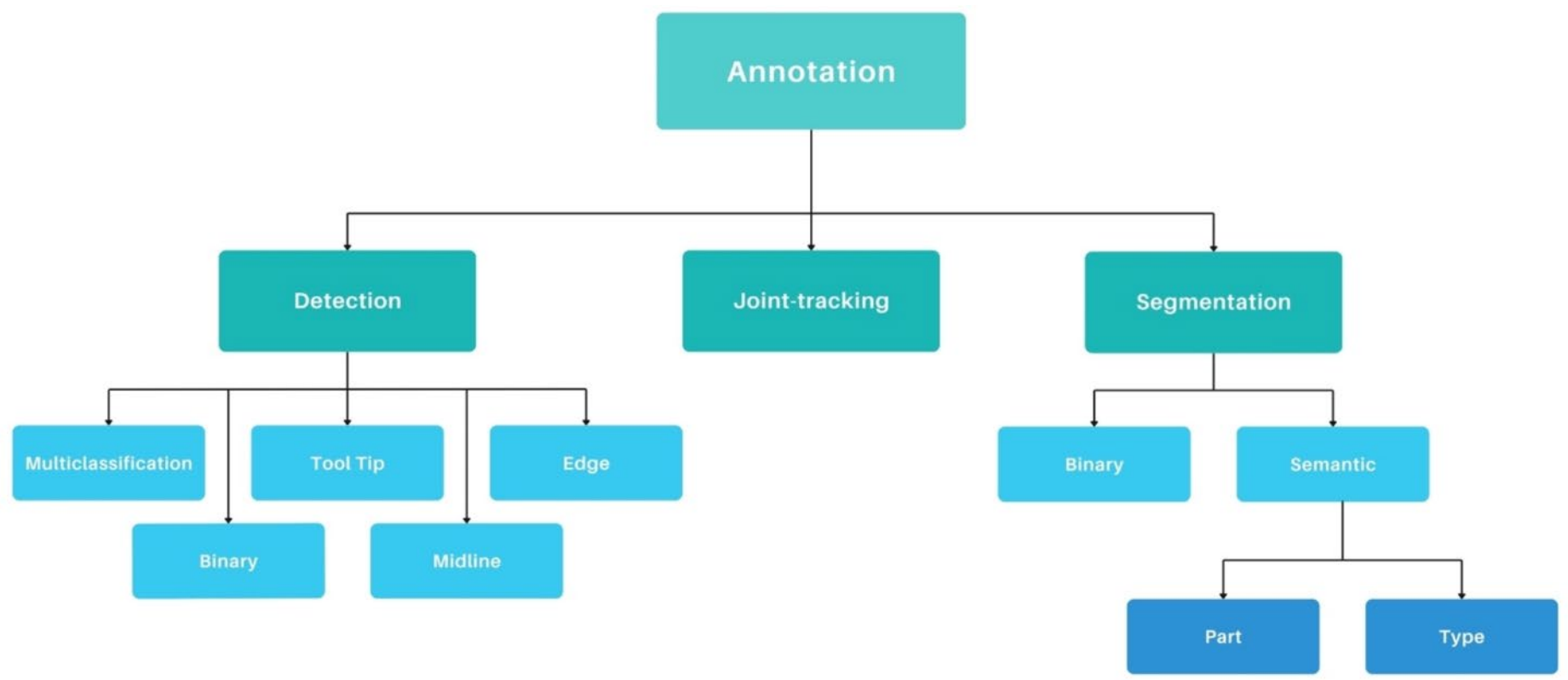

**Fig. 6** Different computer vision techniques for annotation

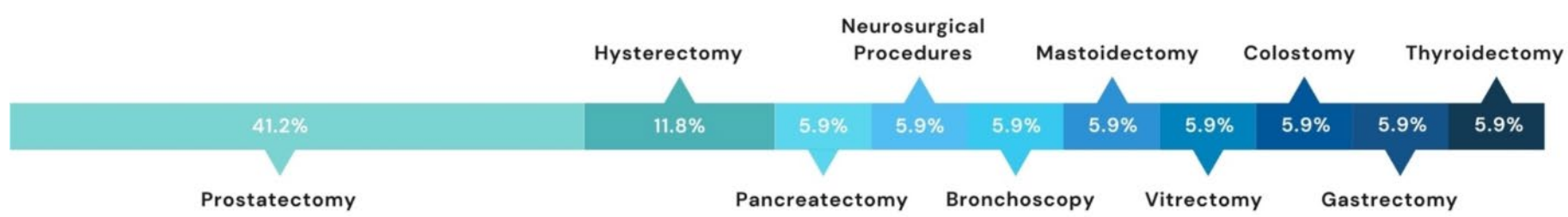

**Fig. 7** Different surgical procedures used for training the developed deep learning models in the included studies

Fig. 7. These datasets included both public and private datasets, which have been further characterized in the sub-sections below.

**Public datasets** Five publicly available datasets were utilized across multiple studies, each focusing on different aspects of surgical instrument detection and segmentation. These datasets include EndoVis2017 and EndoVis2018, which offer comprehensive segmentation tasks for da Vinci surgical instruments. The ARAS-EYE dataset is specific to vitreo-retinal eye surgery, while the RoboTool dataset comprises images from various surgical proce-dures. The Multi-scenery Surgical Tool PUMCH dataset provides annotated endoscopic images from multiple surgeries. Detailed information about each dataset, including the type of procedures, annotations, and dataset size, is summarized in Table 5 below.

**Private datasets** Based on our review of the included 48 articles, 10 utilized their own datasets that were derived from different surgical procedures. These are as follows:

1. Robotic bronchoscopy: Includes 6 classes for tool-presence detection (Background, REBUS, Sheath, Forceps, Needle, and Brush), and 4 classes for episode recognition (Background, REBUS, Needle, and Forceps), developed using the MONARCH® Platform (Zheng et al. 2022).
2. Gynecologic myomectomy and hysterectomy: Comprises 333 video frames manually segmented, yielding 561 segmentation masks for distinct instruments (Kletz et al. 2019).
3. Robotic rectopexy: Includes 49 videos performed by colorectal surgeons and resident trainees, collected using laparoscopic towers and recording systems (J. H. Yang et al. 2022).
4. Radical prostatectomy: Consists of 1,327 frames from 5 radical prostatectomy videos performed using the da Vinci Si surgical system (Kalia et al. 2021).
5. Crowdsourced annotations: Involves key-point annotations for 12 videos (146,309 frames) with a cost of $0.12 per job, (Law et al. 2017). Over 76% of tip regular annotations are within 20 pixels of ground truth annotations. Over 73% of apex regular

**Table 5** Publicly available datasets that were used in the included studies

| Name | Year | Procedure | Annotation | Size |
|---|---|---|---|---|
| Endovis2017, (Allan et al. 2017) | 2017 | abdominal porcine procedures | Binary instrument segmentation, instrument part segmentation, segment and classify the instruments | 10 sequences |
| EndoVis2018, (Allan et al. 2018) | 2018 | Abdominal porcine procedures | Binary instrument segmentation, instrument part segmentation, segment and classify the instruments | 19 sequences |
| RoboTool, (Garcia-Peraza-Herrera et al. 2021) | 2021 | Various freely available surgical procedures on the Internet | Instrument segmentation | 514 images |
| ARAS-EYE dataset, (F. Lotfi et al. 2020) | 2020 | Vitreo-retinal eye surgery | Instrument detection and parts using bounding box | 594 images |
| Multi-scenery Surgical Tool PUMCH, (Ping et al. 2023) | 2023 | Pancreatic, thyroid, colon, gastric surgeries and external scenes | Surgical tools and tool tips detection using bounding box | 181 videos |

annotations are within 25 pixels of ground truth annotations. Under 37% of both annotations were within 25 pixels of ground truth.

## 3.2 Deep learning models

The 48 papers employed various DL models, yet there are variations in the algorithms and applications applied. Notably, CNNs (convolutional neural networks) were the most widely employed methodology and were used either independently or in conjunction with other methodologies. In total, CNNs were used 40 times, vision transformers in 8 models, and GANs in 5 models. Figure 8 shows the distribution of the different deep learning algorithms across the included studies. CNN was the most utilized DL algorithm, seen in nearly 80% of the studies. Within CNN, ResNet (28.6%) and U-Net (26.2%) were most utilized.

### 3.2.1 Hybrid architecture

Many studies integrated two technologies for better performance, such as using ResNet with other architectures like DeeplabV3+ (Yang et al. 2022), U-Net (Xia et al. 2023),Yolov3 (Zinchenko and Song 2021). ResNet was mainly used for feature extraction. U-Net was also used with other different approaches like GANs (Ross et al. 2018). (De Backer et al. 2023) utilized U-Net as a decoder in their proposed network architecture along with Effecient-NetB5 as the encoder. (Hayoz et al. 2023) have reported combining Deeplabv3 + with U-net for binary segmentation for pose estimation.

### 3.2.2 Transfer learning networks

Many articles heavily relied on Transfer Learning, using pre-trained models like U-net and ResNet with different versions such as ResNet18, ResNet34, ResNet50, and ResNet101 and fine-tuned for specific annotation task. U-net was used in 13 studies as a backbone or decoder, while ResNet was used in 12 studies as a backbone as well or feature extractor. Other than CNN pre-trained models, GANs models like CycleGAN with modified network architecture were utilized in different studies to align with the desired goal (Sestini et al.

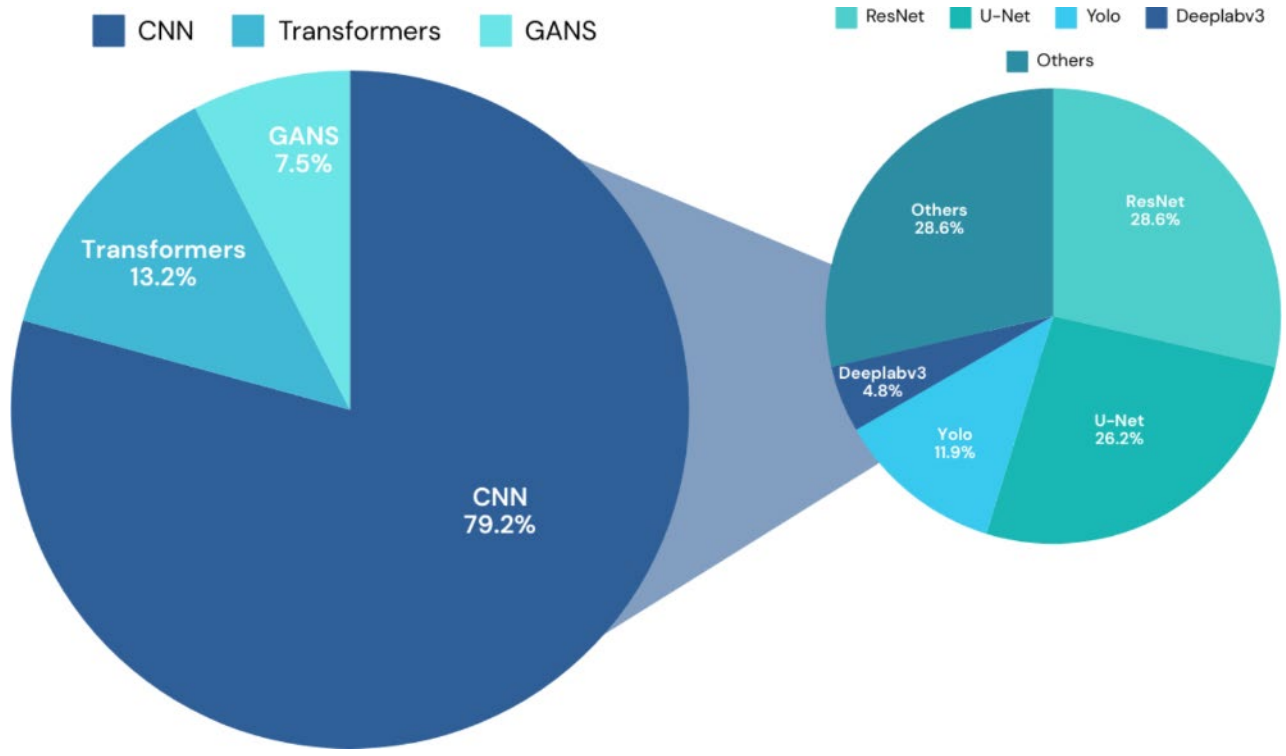

**Fig. 8** Distribution of the different deep learning algorithms across the included studies. The sub-distribution of CNN as the most utilized algorithm is also shown to the right

2022) and (Leifman et al. 2022). YOLO networks were used for real-time object detection in 3 studies (Zinchenko and Song 2021), (Choi et al. 2021), and (Ping et al. 2023). Deeplabv3 and EfficientNetB5 were used as well for transfer learning in few studies either as a backbone or encoder (Wang et al. 2023b, c), (Hayoz et al. 2023), (Kugener et al. 2022), and (De Backer et al. 2023). (Wang et al. 2023b, c) used YOLOv5 for object detection with ResNet18 for feature extraction.

### 3.2.3 Other networks

The remaining papers have utilized other architectures such as vision transformers with ResNet34 as feature extractor (Zheng et al. 2022). (Yang et al. 2022) employed Mask R-CNN to perform tool instance segmentation. It utilized ResNet and feature pyramid network (FPN) as the backbone for feature extraction, and a region proposal network (RPN) to generate object bounding box proposals.

### 3.2.4 Training metrics

A total of 15 studies documented various metrics, including the number of epochs, batch size, learning rate, optimizer, and loss function. The learning rate, commonly set to $10^{-3}$, was linked with optimizers such as Adam, SGD, and Adadelta. Adam was the most utilized optimizer, enhancing model performance in 25 studies. Loss functions documented in 35 studies included cross-entropy, focal loss, Jaccard index, and combined loss functions. Batch sizes, reported in 31 studies, ranged from 2 to 150,000, with a mean size of 8. 28 studies reported a wide range of epochs, ranging from 4 to 2,800, with a mean of approximately 80 (Marullo et al. 2023).

### 3.2.5 Hardware

NVIDIA GPUs were extensively used across the 48 studies, exemplifying the computational demands of DL algorithms. The developed models in the selected studies utilized NVIDIA GPUs ranging from the GeForce GTX series to the Tesla and Quadro series. For training the DL models, six studies used GeForce GTX series which includes GTX 1070 (Huang et al. 2022a), GTX Titan (Du et al. 2018), GTX 3090 (Xia et al. 2023), GTX 1080Ti (Islam et al. 2019), (Colleoni et al. 2019), and (Hasan and Linte 2019). RTX series were used by 12 studies, including RTX-2080ti (Baek et al. 2019) and two RTX 3090 for a complex model that uses STswinCL as framework that integrates transformer with a joint space-time window shift scheme for capturing intra-video relations, (Jin et al. 2022).

### 3.2.6 Performance metrics & scores

A total of 43 studies have documented the performance metrics of the model along with their corresponding scores. The most used performance metrics were mean IoU (Intersection over Union; reported by 13 studies) and DiCE (Diverse Counterfactual Explanations; reported by 10 studies), along with their respective mean and average values. The binary segmentation model achieved a maximum DiCE value of 97.10% and an IoU of 94.4% (De Backer et al. 2023). For IoU, the highest result observed for binary segmentation was 96% (Colleoni and Stoyanov 2021). Other evaluation metrics included accuracy, precision, recall, and mean Average Precision (mAP), with notable tool classification outcomes of 91.53% accuracy, 86.62% precision, and 87.07% recall.

### 3.3 Clinical applications

Out of the 48 included studies, 32 reported clinical applications associated with surgical tool delineation. Figure 9 below presents the different clinical applications that utilize DL in automatic annotation for surgical tools. These include skill assessments (14 studies), post-operative outcome analysis (10 studies), training (6 studies), decision making (6 studies), and surgical workflow analysis (5 studies). Other applications, such as 'surgeon awareness', 'surgical navigation,' 'surgical task automation' and 'surgical report generation' were less commonly mentioned, with only 3 articles mentioning them.

The remaining 16 papers did not declare any clinical applications for the discussed technical work. This distribution of applications highlights the multifaceted impact of DL for surgical tool annotation across different stages of surgical care, from training and intraoperative support to postoperative analysis. For each of these clinical applications, Table 6 below shows the included papers as well as details the deep learning models used, and annotation types utilized, as well as the advantages of obtaining these applications to the practice of surgery.

## 4 Discussion

This systematic review presents an overview of the several DL techniques employed in the detection and segmentation of surgical instruments. A total of 48 studies have implemented DL models for various forms of annotation in robot-assisted surgical videos, including binary segmentation, multi-class segmentation, and instrument components segmentation.

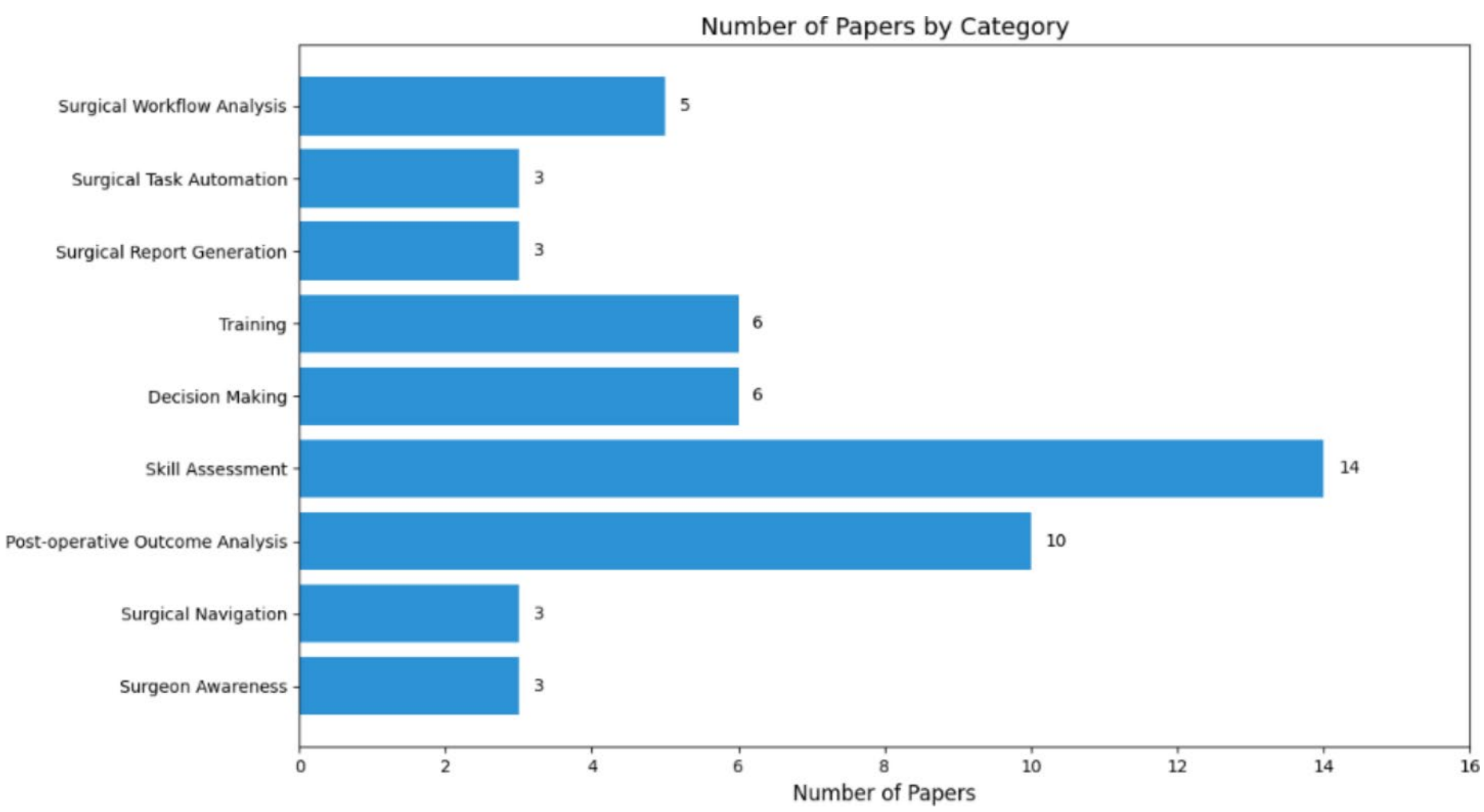

**Fig. 9** Reported clinical applications of surgical tool annotation using deep learning

**Table 6** Clinical applications reported by the included studies for deep learning-based tool annotations

| Clinical applications | Papers | Used annotation | Advantages | Deep learning models |
|---|---|---|---|---|
| Improve Surgeon Awareness | (Huang et al. 2022b), (Tukra et al. 2022), (Xia et al. 2023) | Binary, Parts, and type segmentation | Enhanced situational awareness during surgeries | CNNs, U-Net, GANs |
| Post-operation Outcomes Analysis | (Brandenburg et al. 2023), (Jin et al. 2022), (Kugener et al. 2022), (Law et al. 2017), (Leifman et al. 2022), (Marullo et al. 2023), (Ross et al. 2018), (H. Wang et al. 2023b, c), (Zheng et al. 2022) | Semantic segmentation, Surgical-tool joint detection, Instrument detection | Improved accuracy in outcome predictions | CNNs, ResNet, DeepLab |
| Skill Assessments | (Ping et al. 2023), (Kletz et al. 2019), (Xia et al. 2023), (Choi J, Cho S, Chung JW, Kim, 2021), (Law et al. 2017), (Leifman et al. 2022), (Jin et al. 2022), (Ni et al. 2020), (Garcia-Peraza-Herrera et al. 2021), (Sestini et al. 2022), (Nema and Vachhani, 2023), (Kugener et al. 2022), (Wang et al. 2023b, c), (J. H. Yang et al. 2022), (Li et al. 2023), (Colleoni et al. 2019) | Binary, Parts and type segmentation. Tool and tool tip recognition. Semantic segmentation. Surgical-tool joint detection | Objective measurement of surgical skills for medical school residents | CNNs, ResNet, U-Net, transformers |
| Surgical Navigation | (Jin et al. 2022), (Zinchenko and Song, 2021) | real-time surgical instrument segmentation, semantic segmentation | Precision guidance for surgical instruments | CNNs, U-Net, DeepLab |
| Augmented Reality | (De Backer et al. 2023), (Xia et al. 2023) | Binary segmentation, Parts segmentation, type segmentation | Augmented reality overlays to assist in surgeon training or real time surgeries | CNNs, U-Net, GANs |
| Patient Safety | (Leifman et al. 2022), (Suzuki et al. 2019), (Tukra et al. 2022), (H. Wang et al. 2023b, c) | Semantic segmentation | Enhanced monitoring and prevention of adverse events | CNNs, U-Net, DeepLab |
| Decision Making | (Islam et al. 2019), (Jin et al. 2022), (Li et al. 2023), (Reiter 2022), (Suzuki et al. 2019), (H. Wang et al. 2023b, c) | Tool detection, Binary, parts, instrument segmentation. Semantic segmentation. | Better surgical decisions based on real-time data | CNNs, ResNet, U-Net, transformers |
| Training | (Colleoni et al. 2019), (Leifman et al. 2022), (F. Lotfi et al. 2020), (Ping et al. 2023) | Tool and tool tip recognition, Surgical-tool joint detection and semantic segmentation | Improved training with real-time feedback | CNNs, ResNet, U-Net, GANs, transformers |
| Surgical Task Automation | (Colleoni and Stoyanov 2021), (Garcia-Peraza-Herrera et al. 2021), (Ni et al. 2020) | Real-time segmentation and semantic segmentation | Automation of repetitive or predictable tasks | CNNs, U-Net, DeepLab, GANs |
| Workflow Analysis | (Ayobi et al. 2023), (Leifman et al. 2022), (Ni et al. 2020) | Real-time segmentation and semantic segmentation | Efficient management of surgical workflow | CNNs, U-Net, DeepLab, transformers |

**Table 6** (continued)

| Clinical applications | Papers | Used annotation | Advantages | Deep learning models |
|---|---|---|---|---|
| Report Generation | (Ni et al. 2020), (H. Wang et al. 2023b, c) | Real-time segmentation of robotic surgical instruments | Automated and precise reporting of surgical procedures | CNNs, U-Net, transformers |

It is notable that multi-class and components segmentation cannot be performed without first recognizing the instrument from other non-organic objects within the video frame, i.e. binary segmentation. All the papers included in this analysis focus on the detection of surgical tools using DL models in the context of robotic surgery for in-vivo procedures. The included studies have demonstrated how delineation of articulated instruments is a fundamental block for assorted clinical applications. This review illustrates the potential for incorporation of DL in surgical tool detection, to improve the process of skill assessment, patient safety, post-operation outcome analysis and other diverse clinical needs.

In our review, we found that the diversity of approaches and architectures employed reflects the growing sophistication and versatility of DL models in addressing various challenges within surgical contexts. As we transition to a deeper analysis of specific DL architectures and their performance, it is essential to explore the unique strengths and limitations of these models in different clinical scenarios. Our targeted readership, inclusive of clinicians and computer scientists, will find it equally useful to delve into the comparative performance of key DL architectures, such as U-Net, ResNet, and Transformers. In the following sub-sections, we hope to further discuss the broader implications of DL in surgical applications, including its impact on training, intraoperative procedures, and postoperative analysis.

### 4.1 U-Net and ResNet

U-Net and ResNet are widely recognized for their robust performance in image segmentation tasks, including surgical tool detection and segmentation. Our systematic review includes multiple studies that utilize these architectures and report high accuracy metrics. U-Net is specifically designed for biomedical image segmentation and has shown exceptional performance due to its encoder-decoder structure with skip connections. This structural design allows for precise localization and segmentation of surgical tools while preserving contextual information by fusing low-level characteristics from the encoder with high-level features from the decoder. For example, (Huang et al. 2022b) reported a Dice coefficient of 0.945 and an IoU of 0.883 using a U-Net model enhanced with a morphological polar transform. ResNet is the state-of-the-art network in feature extraction, making it a suitable choice as an encoder in DL models. ResNet's strength lies in its deep residual learning framework, which mitigates the vanishing gradient problem in deep networks using residual blocks, which include skip connections which link activations to subsequent layers. ResNet's flexibility, offered in various versions like ResNet18, ResNet34, and ResNet101, makes it adaptable for both small and large datasets. Studies such as (Ni et al. 2020) have demonstrated its effectiveness and adaptability. (Ni et al. 2020) reported an mIoU of 94.10% and an mDice of 96.91% using an attention-guided lightweight network based on ResNet.

Compared to other architectures, U-Net and ResNet often outperform due to their unique structural advantages. For instance, (Cai and Zhao 2020) used a two three-layer CNN frame-work and reported an accuracy of 75% on the EndoVis dataset, which is notably lower U-Net and ResNet models in similar contexts. Additionally, (Colleoni et al. 2019) utilized an encoder-decoder architecture with 3D convolutions and achieved a Dice similarity coefficient of 85.1% for joint detection, also lower than U-Net and ResNet. Based on comparative analysis, U-Net and ResNet exhibit superior accuracy in surgical tool segmentation tasks compared to other architectures. U-Net's effective feature preservation and localization capabilities make it ideal for high-precision tasks, while ResNet's ability to train deeper networks without degradation is advantageous in complex pattern recognition scenarios. In specific surgical scenarios, U-Net excels in tasks requiring high spatial accuracy, such as tumor boundary detection, while ResNet's deeper architecture is better suited for complex hierarchical feature recognition, such as differentiating overlapping instruments in robotic surgeries. Future research should focus on optimizing these architectures for specific surgical applications to further enhance their performance.

### 4.2 Transformers

Transformers are seldom utilized in network architectures; however, three studies (Jin et al. 2022), (F. Wang et al. 2023b, c), and (Xu et al. 2022) used Swin transformer in addition to CNNs. By integrating Swin transformers with CNNs, the DL model may effectively employ labels or pseudo labels to improve the accuracy of pair generation in instrument segmentation. However, their primary emphasis is entirely on the semantic segmentation of a single picture (Jin et al. 2022). Masked-Attention Transformers for Instrument Segmentation is a transformer-based method that uses masked and deformable attention to segment instrument instances. It enhances mask classification using video transformers. Mask2Former is MATIS' instance segmentation baseline, which utilizes a Swin Transformer backbone. It incorporates a multi-scale deformable attention pixel decoder and masked attention algorithms (Ayobi et al. 2023). (Xu et al. 2022) used transformers to develop an end-to-end detector and feature extractor-free captioning model using the patch-based shifting window approach. This design obviates the need of using a feature extractor, such as CNNs, as transformers are intricate models that require substantial processing resources. We also noticed that the papers deploying transformers used the most advanced hardware, as they consume huge computational power. (Ayobi et al. 2023) used 4 NVIDIA Quadro RTX 8000 GPUs for the masked attention baseline and a single NVIDIA Quadro RTX 8000 GPU for models requiring substantial computational resources. Most studies performed transfer learning which does not need an extensive computing power as most layers are pre-trained, with only a few layers requiring actual training.

### 4.3 GANs and the creation of synthetic data

One of the limitations pointed out in several papers is the lack of data; therefore, the creation of synthetic data is an optimal solution. The use of surgical instrument annotation for creating a dataset indistinguishable from real surgical procedures is crucial. Instrument detection is essential, as the DL model should be able to extract all the necessary features and objects within a frame to replicate another procedure. This was accomplished by using GANs, where the network consists of an encoder for feature extraction and a decoder for the creation of the new frame, (Colleoni and Stoyanov 2021). Another use of synthetic data is incorporating it into the training datasets to have more examples and testing for better model performance. CycleGAN, a pretrained GAN network, was used for such a task along with other software for data construction, like Blender 3D (Leifman et al. 2022).

## 4.4 Hyperparameters and performance scores

The included studies in this systematic review have reported multiple hyperparameters that control the model's performance. These metrics include the number of epochs, batch size, loss function, learning rate, and optimizer. The most important hyperparameter in any DL model is the loss function, which measures the difference between the predicted output and the ground truth. The goal in any DL model is to minimize the value of the loss function as much as possible for better model performance. This is done through multiple iterations and the utilization of an optimizer. Depending on the task performed, a certain loss function would be optimal. Notably, papers performing binary segmentation used either binary cross-entropy, focal loss, or DiCE loss. On the other hand, studies deploying multiclass segmentation utilized categorical cross-entropy. GAN models used adversarial loss (Colleoni and Stoyanov 2021), (Nema and Vachhani, 2023), (Tukra et al. 2022), or combination of reconstruction loss, perceptual loss, style loss, warping loss, and total variation loss (Kalia et al. 2021). Studies developing models based on autoencoders used mean squared error loss or binary cross-entropy loss. Finally, as all papers are performing object detection, most of the papers used cross-entropy. Other papers that used multiple models used IoU loss along with other loss functions, such as BCEWithLogits loss (Xia et al. 2023). This is most likely due to these papers using multiple models for different purposes, for example YOLOv5 for object detection, ResNet18 for feature extraction, and node tracking mechanism, and the M2 transformer for surgical report generation (H. Wang et al. 2023b, c).

## 4.5 Tool detection for skill assessments and training

The incorporation of DL into surgical training enhances the precision of tool detection and segmentation, offering an in-depth analysis of surgical tool dynamics and interactions. This signifies a critical evolution in surgical training methods. For example, instance segmentation technologies, as highlighted in (Wang et al. 2023b, c), enable accurate identification and monitoring of individual surgical instruments within complex operational scenarios. This feature is vital for evaluating tool positioning and manipulation, which are crucial indicators of a surgeon's expertise. Real-time semantic segmentation, as evidenced (Law et al. 2017), provides instant feedback on tool handling, fostering a dynamic evaluation environment. These advancements are particularly beneficial in training settings, where an immediate understanding of tool-tissue interactions can significantly elevate a novice surgeon's learning experience. Moreover, technologies like tool-tip detection and multi-class segmentation provide detailed insights into specific tool handling aspects (Ping et al. 2023).

*Impact on Surgery and Surgical training*: Integrating these technological solutions into simulated training setups marks a transformative advancement. Utilizing datasets such as 'EndoVis2017' and tailored datasets from specific surgeries, training initiatives can simulate a range of surgical scenarios that mimic real-life complexities but without the inherent risks. For example, the 'Multi-scenery Surgical Tool PUMCH' dataset, which includes varied surgical environments, offers extensive visual and contextual diversity, thus equipping trainees for numerous surgical challenges (Ping et al. 2023). Additionally, the progression towards automated and semi-automated annotation techniques simplifies and democratizes the training process. Methods like HSV thresholding and GrabCut in the 'RoboTool' dataset lessen reliance on expertly labeled data, often a major constraint in creating training materials (Garcia-Peraza-Herrera et al. 2021). This shift not only broadens the scalability of training programs but also ensures consistent training data quality, essential for upholding educational excellence (Ni et al. 2020).

The implications of these technological advancements in surgery are profound. Enhanced training tools lead to better-equipped surgeons, directly influencing improved patient outcomes (Ross et al. 2018). The ability to standardize training using scalable DL technologies across various regions and institutions can help reduce disparities in the quality of surgical care. Additionally, the immediate feedback provided by these technologies shortens the learning curve for surgical trainees, enabling them to master complex techniques more swiftly and confidently (Ayobi et al. 2023).

### 4.6 Post-surgical applications

Our review also highlights the integral role of DL in postoperative settings, particularly through meticulous segmentation and analysis of surgical tools captured in surgical video footage. These precise segmentation capabilities are seen in studies using datasets like 'EndoVis2017' and 'RoboTool' that enable detailed postoperative reviews where surgical maneuvers are closely examined (Leifman et al. 2022). These segmentation techniques distinguish between different tool types and their interactions with the surgical field, offering an in-depth look at the procedural nuances. Such detailed observation is essential for pinpointing critical surgical moments that might influence patient outcomes. For instance, research demonstrated in studies (Cai and Zhao 2020) and (Ping et al. 2023) show that real-time semantic segmentation can retrospectively identify and scrutinize pivotal surgical phases where the handling of tools may be linked to either complications or successes. This retrospective analysis helps surgical teams understand specific actions that might be improved or adjusted in subsequent procedures. Additionally, the application of DL in postoperative reviews aids in the continual enhancement of surgical methods (Brandenburg et al. 2023). Through the analysis of outcomes from various surgeries, enabled by DL-powered video analytics, patterns that lead to superior outcomes can be discerned (Zheng et al. 2022). This not only aids in the professional development of individual surgeons but also contributes to the broader scope of surgical training and protocol refinement.

*Impact on Surgery*: The impact of these technological advancements extends beyond individual outcomes, improving overall healthcare quality. The detailed data provided by these technologies support healthcare facilities in auditing and standardizing surgical practices, ensuring adherence to stringent safety and efficiency guidelines (Tukra et al. 2022). This improvement in procedural consistency bolsters patient safety and trust in surgical care. Moreover, the employment of automated and semi-automated tool annotations minimizes human error in postoperative analysis and enhances the efficiency of these evaluations (Wang et al. 2023b, c). This leads to more uniform and thorough audits, which are crucial for upholding high care standards and promoting ongoing enhancement in surgical practices (Brandenburg et al. 2023).

The broader implications of these postoperative applications in surgery are significant. Advanced tool detection and segmentation technologies foster a deeper understanding of surgical procedure intricacies, which directly impacts training programs, protocol formulation, and ultimately, standards of patient care. By refining the scope and accuracy of post-operative reviews, these technologies enable surgical teams to more effectively identify and address risks, leading to improved patient outcomes (Colleoni and Stoyanov 2021). Furthermore, the systematic collection of segmented surgical data supports extensive studies aimed at enhancing surgical techniques and outcomes across various surgeries and patient groups (Suzuki et al. 2019). They equip the surgical community with essential tools to increase the precision, safety, and efficacy of surgical operations, promoting a culture of continuous learning and advancement that is crucial to contemporary medical practice.

### 4.7 Intra-surgical applications

DL applications within the intraoperative phase substantially enhance surgical precision by utilizing real-time tool detection and segmentation. The employment of CNNs and other DL models, as highlighted in studies using datasets like 'EndoVis2017' and 'RoboTool', facilitates the immediate identification and categorization of various surgical instruments during operations (Zheng et al. 2022), (Colleoni and Stoyanov 2021), (Ni et al. 2020), and (Garcia-Peraza-Herrera et al. 2021). This capability is crucial for maintaining situational awareness, especially during complex and minimally invasive surgeries where visibility and access may be limited The precise recognition and segmentation of different surgical tool components, such as those emphasized in studies (Huang et al. 2022a) and (Tukra et al. 2022) focusing on tool-tip detection, are vital in aiding surgeons to execute meticulous movements intraoperatively. It could offer visual aids and data that assist in navigating the surgical field, thus reducing the cognitive burden on surgeons. Lowering this cognitive load is essential as it enables surgeons to concentrate more on crucial decision-making processes and less on the intricacies of tool manipulation, potentially reducing surgical mistakes (Xia et al. 2023). Additionally, integrating these DL technologies with robotic systems like the da Vinci surgical platforms enhances the interaction between surgeons and robotic tools. For instance, real-time semantic segmentation can be aligned with robotic arm movements to continuously optimize tool positioning and manipulation during surgeries (De Backer et al. 2023). This integration facilitates smoother procedural flows and augments the capabilities of robotic surgeries, increasing their efficiency and reducing susceptibility to human error.

*Impact on Surgery*: The implications of these intraoperative applications are profound in the surgical field. They signify a shift in surgical procedures, particularly with the integration of cutting-edge technologies and human expertise. By improving the precision and efficiency of surgeries, these technologies can make significant contributions to better patient outcomes and quicker recovery periods (Law et al. 2017). Moreover, the intraoperative support provided by DL technologies is crucial for training surgeons on robotic platforms (F. Lotfi et al. 2020). The comprehensive feedback and data provided by these systems help trainees understand the dynamics of robotic tools and their application in various surgical contexts (Leifman et al. 2022). This training is invaluable as it equips surgeons to manage the complexities associated with the increasing prevalence of robotic surgeries in contemporary healthcare. The application of these technologies also promotes a collaborative environment where technological innovation and human expertise merge to extend the possibilities of surgical achievements (H. Wang et al. 2023b, c). This synergy not only improves the surgical process itself but also accelerates the development of new surgical techniques and innovations.

Additionally, surgical tool segmentation can contribute to the development of augmented reality (AR). Remarkably, AR can assist surgeons in executing accurate surgical procedures (De Backer et al. 2023). One example of how AR might enhance surgeons' visual perception of high-risk targets is through the use of endoscopic footage (Xia et al. 2023). All these applications contribute to patient safety and the smooth performance of the procedure. Notably, with the variety of different applications, multiple DL models would be employed to achieve such tasks, like different CNNs and transducers. These advanced technologies can also aid in reducing the risk of human error during surgeries. By providing real-time feedback and guidance, AR can enhance the precision and efficiency of surgical interventions.

### 4.8 Impact on surgical team dynamics and communication

The integration of AI and DL technologies into surgical environments not only enhances the precision and efficiency of procedures but also influences team dynamics and communication. As discussed in Sect. 4.5, the adoption of DL tools in surgical training has significantly improved the real-time detection and segmentation of tools, which contributes to more effective and coordinated team operations (Colleoni et al. 2019). These advancements promote a more synchronized workflow by providing real-time data and visualizations that are accessible to all team members, reducing the reliance on verbal communication and minimizing the risk of misunderstandings (Huang et al. 2022a).

Moreover, as these technologies become increasingly embedded in surgical practice, there is an emerging need for surgical teams to adapt to new workflows and interaction patterns. This requires a comprehensive understanding of the capabilities and limitations of DL tools, which could be achieved through targeted holistic training programs. A critical component of such programs, as noted in Sect. 4.5, should include technical training on commonly used DL models like U-Net and ResNet. essential for understanding the functioning of AI tools to help team members interpret the data and visualizations these tools provide. In addition to technical training, simulation-based training would allow for hands-on practice in a controlled environment, helping the team become familiar with new workflows and communication patterns without the pressures of a real surgical scenario (Ping et al. 2023). Team coordination exercises are also important, focusing on improving communication and decision-making processes within the team when using DL tools (Li et al. 2023). As AI technologies continue to evolve, continuous education would become necessary to keep the teams updated on the latest advancements, through regular workshops, seminars, and online courses. Finally, ethical and safety training is crucial, ensuring that the team understands the ethical implications, data privacy concerns, and appropriate use of AI-generated data to maintain patient safety.

### 4.9 Ethical concerns and data privacy

The use of surgical data for training DL models raises several ethical concerns, primarily related to patient privacy and data security. Ensuring the confidentiality of patient information is paramount when dealing with sensitive medical data. These could be addressed in a couple of ways: obtaining informed consent and surgical data anonymization. Obtaining informed consent from patients is a crucial first step in any medical data collection protocol. Patients should be fully informed about how their data will be used, including the specific purposes of the research and any potential risks involved. Without informed consent, the use of patient data would be unethical and could lead to significant privacy issues (Arora and Thota 2024). Secondly, surgical data must be thoroughly anonymized to remove any identifiable information. This involves not only stripping direct identifiers such as names and medical record numbers but also indirect identifiers that could potentially be used to trace back to the patient (Murdoch 2021). Advanced anonymization techniques, including de-identification and pseudonymization, are essential to protect patient privacy. De-identification involves removing all identifiable information from the dataset, while pseudonymization replaces private identifiers with fictitious names or codes (Yoon et al. 2020).

Data security is another critical aspect that needs to be considered while training DL (or any AI) models. Ensuring the secure storage and transmission of data is essential to prevent unauthorized access. This includes the use of encryption and secure protocols for data handling. Implementing strict access control measures, such as multi-factor authentication and role-based access controls, ensures that only authorized personnel have access to the data (Kaissis et al. 2021). Importantly, any research involving patient data should undergo ethical review by an Institutional Review Board (IRB) or equivalent ethics committee (Amdur and Biddle 1997). This review process ensures that the research complies with ethical standards and regulations, providing an additional layer of oversight. Finally, when sharing large, annotated datasets, establishing formal data sharing agreements is important. These agreements should clearly delineate the responsibilities and limitations of data use, stipulating the conditions under which the data can be used and ensuring compliance with privacy laws and ethical guidelines (Batlle et al. 2021). Transparency with patients and the public about the use of surgical data in research is also crucial. Public disclosures about the types of data being collected and the purposes for which it is used can help foster trust and accountability (Andreotta et al. 2022).

### 4.10 Limitations

The systematic review presents the challenges that are associated with the use of DL techniques in robotic surgery, particularly for tasks such as segmenting, detecting, and accurately recognizing surgical instruments. The most frequent dilemma is the limited number of testing videos that are accessible for model validation. This limitation arises from the difficulties involved in obtaining a sufficiently extensive and varied set of actual surgical video footage. This challenge represents a larger problem in medical image analysis, where the limited availability of comprehensive and diverse datasets hinders the progress, evaluation, and validation of sophisticated DL models. Most of the used data is from porcine models, indicating a deficiency in the number of human data and real annotated data needed for better models' development.

### 4.11 Future directions

Future research should focus on the creation and sharing of large, annotated datasets from diverse surgical procedures to address this limitation. Collaborative efforts across institutions to compile and annotate such datasets could significantly enhance the training and vali-dation of DL models. Additionally, exploring the integration of synthetic data to supplement real-world data could help overcome the data scarcity issue. There is also a need for more robust models that can generalize well across different surgical environments and lighting conditions. Research should continue to refine these models, incorporating advancements in DL architectures, such as transformers and GANs, to improve their performance and applicability in surgical settings.

### 4.12 Conclusion

The application of DL in annotating surgical instruments holds immense promise for enhancing surgical precision, training, and outcomes. This systematic review has thoroughly examined the effectiveness of DL in the labeling of surgical equipment, demonstrating significant improvements in the accuracy and speed of these processes. Our investigation reveals that DL, namely using CNNs and advanced architectures like U-Net and ResNet, greatly enhances surgical tool detection and segmentation. This technical innovation is essential for a wide range of applications, including real-time surgical navigation and exhaustive postoperative evaluations, and plays a critical role in enhancing surgical results and ensuring patient safety.

The importance of these discoveries is in the capacity of DL to facilitate intricate medical procedures and training initiatives. Improved precision in identifying and separating tools immediately leads to decreased risks during surgery and enhanced accuracy, both of which are decisive for ensuring patient safety and effective surgical procedures. However, the review highlights important limitations, such as the lack of varied and comprehensive datasets, which could impact the applicability and reliability of the DL models. Additionally, the reliance on well-annotated data of superior quality for successful model training presents further challenges, constraining the ability to scale and use the model in different clinical environments.

Exploring the fusion of DL with AR technology has the potential to advance the creation of more user-friendly and intuitive surgical systems. Furthermore, given the rapid advancement of DL architectures, it is imperative to continuously assess new models in clinical settings. Future research should focus on addressing these limitations by creating and sharing large, annotated datasets from diverse surgical procedures and integrating synthetic data to supplement real-world data.

**Acknowledgements** The authors would like to acknowledge the support of the Surgical Research Section and the Clinical Advancement Department at Hamad Medical Corporation for the conduct of this research.

**Author contributions** SB conceptualized the review and methodology, and supervised the work. FA formulated the research question based on iterative preliminary database searches, defined the scope of the review, and built a comprehensive search strategy under the supervision of SB. FA, MY and AM collaboratively performed title abstract screening. Full-text screening was done by MA, MY, AM and FA, and conflicts were resolved mutually through discussion or under consultation from SB and OA. FA designed the data extraction sheet and labels, and conducted the extraction with MY, MA and HO. Data synthesis was led by FA, with iterative input from HA, ZS and SB. FA led the drafting of the original manuscript, with critical inputs from MA, HA and SB. OA and AA provided iterative input for subsequent drafts. SB, OA and AA led the discussion of surgical applications of the findings. All authors were actively involved in data curation, writing, review, and editing of the manuscript.

**Funding** Open Access funding provided by the Qatar National Library. The research was conducted as part of this project. ARG01-0522-230266 is the award grant number. Research reported in this publication was supported by the Qatar Research Development and Innovation Council (QRDI) grant number ARG01-0522-230266. Open access funding provided by Qatar National Library (QNL). *Disclaimer*: The content is solely the responsibility of the authors and does not necessarily represent the official views of Qatar Research Development and Innovation Council or Qatar National Library.

**Data availability** No datasets were generated or analysed during the current study.

## Declarations

**Conflict of interest** The authors declare that they have no conflict of interest.

**Research involving human participants and/or animals** This article does not contain any studies with human participants or animals performed by any of the authors.

**Informed consent** Not applicable, as this research did not involve human participants.

## Authors and Affiliations

**Fatimaelzahraa Ali Ahmed[1] · Mahmoud Yousef[2] · Mariam Ali Ahmed[3] · Hasan Omar Ali[2] · Anns Mahboob[2] · Hazrat Ali[6] · Zubair Shah[4] · Omar Aboumarzouk[1] · Abdulla Al Ansari[1] · Shidin Balakrishnan[1]**

✉ Shidin Balakrishnan
sbalakrishnan1@hamad.qa

[1] Department of Surgery, Hamad Medical Corporation, Doha, Qatar

[2] Weill Cornell Medicine, Doha, Qatar

[3] College of Medicine, Qatar University, Doha, Qatar

[4] College of Science and Engineering, Hamad Bin Khalifa University, Doha, Qatar

[6] Computing Science and Mathematics, University of Stirling, Stirling, United Kingdom